\documentclass{ws-procs9x6}

\providecommand{\Journal}[4] {#1\textbf{#2}, #3 (#4)}
\providecommand{\CPC}{Comput. Phys. Commun. } %
\providecommand{\NIMA}{Nucl. Instr. Meth. \textbf{A}} %
\providecommand{\NPA}{Nucl. Phys. \textbf{A}} %
\providecommand{\NPB}{Nucl. Phys. \textbf{B}} %

\providecommand{\PAN}{Phys. Atom. Nucl. } %
\providecommand{\PLB}{Phys. Lett. \textbf{B}} %
\providecommand{\PRL}{Phys. Rev. Lett. } %
\providecommand{\PRC}{Phys. Rev. \textbf{C}} %
\providecommand{\PRD}{Phys. Rev. \textbf{D}} %
\providecommand{\YF}{Yad. Fiz. } %
\providecommand{\ZPA}{Z. Phys. \textbf{A}} %
\providecommand{\ibid}{{\it ibid. }} %

\providecommand{\Thp}{$\Theta^+$~}
\providecommand{\pythia}{{\sc Pythia6}}

\begin{document}

\title{Pentaquark Search at HERMES\footnote{\uppercase{T}his work is supported by
the \uppercase{U.S.} \uppercase{N}ational \uppercase{S}cience
\uppercase{F}oundation under Grant 0244842.}}

\author{W.~Lorenzon (\lowercase{on behalf of the} HERMES C\lowercase{ollaboration})}

\address{Randall Laboratory of Physics,\\
University of Michgan,\\
Ann Arbor,  Michigan 48109-1120, USA\\
E-mail: lorenzon@umich.edu}

\maketitle

\abstracts{Evidence for a narrow baryon state at $1528 \pm
2.6\mbox{(stat)} \pm 2.1\mbox{(syst)}$\,MeV is presented in
quasi-real photoproduction on a deuterium target through the decay
channel $p K^0_S \to p \pi^+ \pi^-$. The statistical significance
of the peak in the $p K^0_S $ ~invariant mass spectrum is 4
standard deviations and its extracted intrinsic width $\Gamma =
17\pm9$(stat)$\pm3$(syst)\,MeV. This state may be interpreted as
the predicted $S$=$+1$ exotic $\Theta^{+}(uudd\bar{s})$ pentaquark
baryon.}

\section{Introduction}
One of the central mysteries of hadronic physics has been the
failure to observe baryon states beyond those whose quantum
numbers can be explained in terms of three quark configurations.
Exotic hadrons with manifestly more complex quark structures, in
particular exotics consisting of five quarks, were proposed on the
basis of quark and bag models\cite{Jaff76} in the early days of
QCD.

More recently, an exotic baryon of spin 1/2, isospin 0, and
strangeness $S$=$+1$ was discussed as a feature of the Chiral
Quark Soliton model.\cite{Che85} In this
approach\cite{Wal92,Dia97} the baryons are rotational states of
the soliton nucleon in spin and isospin space, and the lightest
exotic baryon lies at the apex of an anti-decuplet with spin 1/2,
which corresponds to the third rotational excitation in a three
flavor system. Treating the known N(1710) resonance as a member of
the anti-decuplet, Diakanov, Petrov, and Polyakov\cite{Dia97}
derived a mass of 1530\,MeV and a width of less than 15\,MeV for
this exotic baryon, since named the $\Theta^+$. It corresponds to
a $uudd\overline{s}$ configuration, and decays through the
channels $\Theta^{+}\rightarrow pK^0$ or $nK^{+}$. However,
measurements of $K^+$ scattering on proton and deuteron targets
showed no evidence\cite{Arn03} for strange baryon resonances, and
appear to limit the width to remarkably small values of order an
MeV.

Experimental evidence for the $\Theta^{+}$ first came from the
observation of a narrow resonance\cite{SPring8} at $1540\pm
10\mbox{(syst)}$\,MeV in the $K^-$ missing mass spectrum for the
$\gamma n \to K^+K^-n$ reaction on $^{12}$C. This result was
confirmed since then by a series of experiments, with the
observation of sharp peaks\cite{DIANA}\cdash\cite{LPI} in the $nK^+$
and $pK^0_S$ invariant mass spectra near 1530\,MeV, in most cases
with a width limited by the experimental resolution. There are also
many unpublished reports of failures to observe this signal.

\section{Experiment}\label{sec:PID}

Presented here are the results of a search for the $\Theta^+$ in
quasi-real photoproduction on deuterium.\cite{HERMES} The data were
obtained by the HERMES experiment with the 27.6\,GeV positron beam
of the HERA storage ring at DESY. The HERMES spectrometer is
described in detail in Ref.~\refcite{SPE}. The analysis searched for
inclusive photoproduction of the $\Theta^+$ followed by the decay
$\Theta^+\to p K^0_S \to p \pi^+ \pi^-$. Events selected contained
at least three tracks: two oppositely charged pions in coincidence
with one proton. Identification of charged pions and protons was
accomplished with a Ring-Imaging \v{C}erenkov (RICH)
detector,\cite{RICH} which provides separation of pions, kaons and
protons over most of the kinematic acceptance of the spectrometer.
In order to keep the contaminations for pions and protons at
negligible levels, protons were restricted to a momentum range of
4--9\,GeV/c and pions to a range of 1--15\,GeV/c.

The event selection included constraints on the event topology to
maximize the yield of the $K^0_S$ peak in the $M_{\pi^+\pi^-}$
spectrum while minimizing its background. Based on the intrinsic
tracking resolution, the required event topology included a
minimum distance of approach between the two pion tracks less than
1\,cm, a minimum distance of approach between the proton and
reconstructed $K^0_S$ tracks less than 6\,mm, a radial distance of
the production vertex from the positron beam axis less than 4\,mm,
a $z$ coordinate of the production vertex within the $\pm 20$\,cm
long target cell of $-18$\,cm $< z < +18\,$cm along the beam
direction, and a $K^0_S$ decay length greater than 7\,cm.  To
suppress contamination from the $\Lambda(1116)$ hyperon, events
were rejected where the invariant mass $M_{p \pi^-}$ fell within
$2\,\sigma$ of the nominal $\Lambda$ mass, where $\sigma=2.6$\,MeV
is the apparent width of the $\Lambda$ peak observed in this
experiment.

\section{Results}
The resulting invariant $M_{\pi^+\pi^-}$ spectrum yields a $K^0_S$
peak at $496.8 \pm 0.2$\,MeV, which is within 1\,MeV of the expected
value of $497.7 \pm 0.03$\,MeV.\cite{PDG} To search for the
$\Theta^+$, events were selected with a $M_{\pi^+\pi^-}$ invariant
mass within $\pm 2\,\sigma$ about the centroid of the $K_S^0$ peak.
The resulting spectrum of the invariant mass of the $p\pi^+\pi^-$
system is displayed in Fig.~1 (left panel).
\begin{figure}[ht]
\centerline{\epsfxsize=54mm\epsfbox{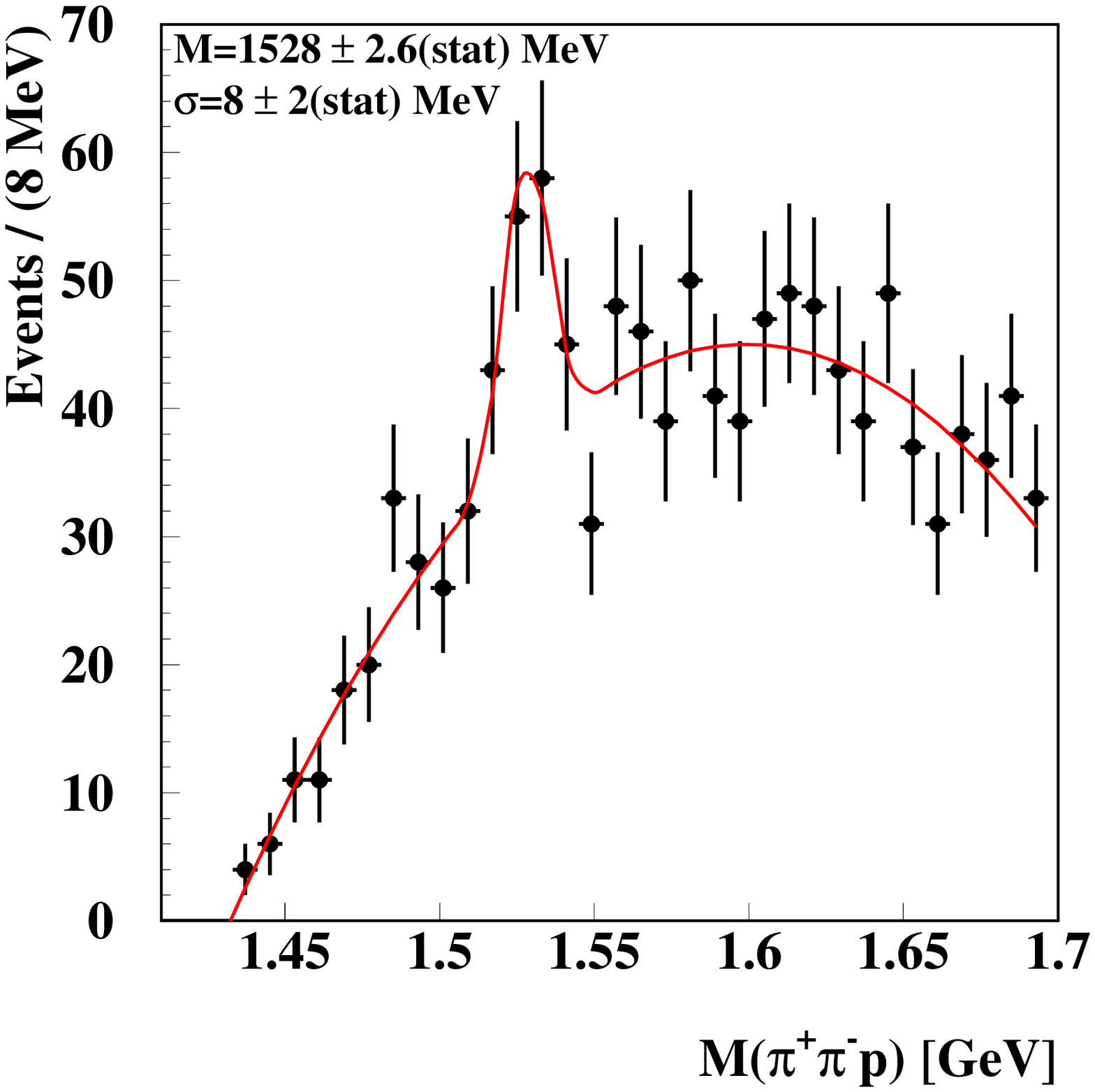}\ \hspace*{2mm} \
\epsfxsize=54mm\epsfbox{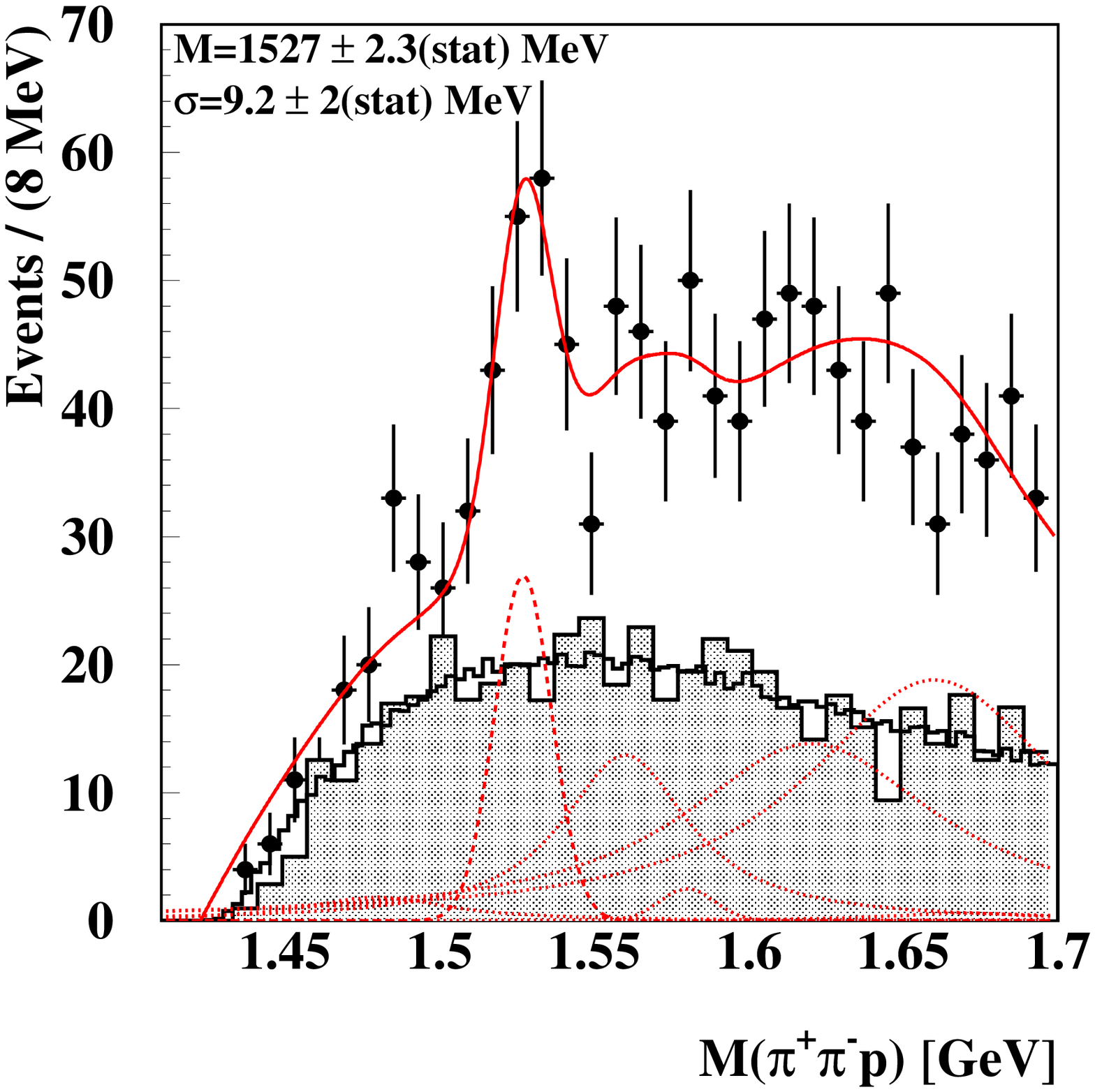}}

\caption{Distribution in invariant mass of the $p \pi^+\pi^-$
system subject to various constraints described in the text.
Experimental data are represented by filled circles with
statistical error bars, while fitted smooth curves result in
indicated position and $\sigma$ width of the peak of interest.
Left panel: a fit to the data of a Gaussian plus a third-order
polynomial is shown. Right panel: the \pythia\ Monte Carlo
simulation is represented by the gray shaded histogram, the
mixed-event model normalized to the \pythia\ simulation by the
fine-binned histogram, the known $\Sigma^{*+}$ resonances by the
dotted curves, and the narrow Gaussian for the peak of interest by
the dashed curve. \label{Fig1}}

\end{figure}

A narrow peak is observed at $1528.0\pm 2.6\pm 2.1$\,MeV with a
Gaussian width of $\sigma=8\pm2$\,MeV and a statistical
significance of $N_{s}/{\delta N_{s}} = 3.7$. Here, $N_s$ is the
full area of the peak from a fit to the data of a Gaussian plus a
third-order polynomial, and $\delta N_s$ is its fully correlated
uncertainty. All correlated uncertainties from the fit, including
those of the background parameters, are accounted for in $\delta
N_s$. There is no known positively charged strangeness-containing
baryon in this mass region (other than the $\Theta^+$) that could
account for the  observed peak.

In an attempt to better understand the signal and its background,
two additional models for the background were
explored.\cite{HERMES} For the first model, a version of the
\pythia\ code\cite{PYTHIA6} tuned for HERMES
kinematics\cite{ourPYTHIA6} is taken to represent the non-resonant
background, and the remaining  strength in the spectrum is
attributed to a combination of known broad resonances and a new
structure near 1.53\,GeV. For the second model, the non-resonant
background is simulated by combining from different events a kaon
and proton that satisfy the same kinematical requirements as the
tracks taken from single events in the main analysis. This
procedure yields a shape that is very similar to that from the
\pythia\ simulation, as shown in Fig.~1 (right panel). By fitting
a polynomial to the mixed-event background normalized to the
\pythia\ simulation, a peak is obtained at $1527.0\pm 2.3\pm
2.1$\,MeV with a Gaussian width of $\sigma=9.2\pm 2$\,MeV and a
statistical significance of $N_{s}/{\delta N_{s}} = 4.3$. The
resulting values from the two fits for the centroid are found to
be consistent, while the width and significance depend on the
method chosen to describe the remaining strength of the spectrum.

Using a \Thp ``toy Monte Carlo'' with
$\Gamma_{\Theta^+}$=$2$\,MeV, an instrumental width of
10--14.6\,MeV (FWHM) was derived. This is somewhat smaller than
the observed 19--24\,MeV (FWHM) width of the peak.\footnote{The
indicated range in width depends on the background model and on
the mass reconstruction method used.\cite{HERMES}} Therefore, the
peak of interest was re-fit with a Breit-Wigner form convoluted
with a Gaussian whose width was fixed at the simulated resolution.
The resulting value for the intrinsic width is $\Gamma =
17\pm9$(stat)$\pm3$(syst)\,MeV.

In order to study the isospin of the observed resonance, the
possibility that the $\Theta^{++}$ partner is present in the $M_{p
K^+}$ spectrum was explored. Although Fig.~2 shows a clear peak
for the $\Lambda(1520)$ in the $M_{pK^-}$ invariant mass spectrum,
there is no peak structure observed in the $M_{pK^+}$ invariant
mass distribution. The upper limit of zero counts is at the 91\%
confidence level. The failure to observe a $\Theta^{++}$ suggests
that the observed $\Theta$ is not isotensor and is probably
isoscalar.

\begin{figure}[ht]
\centerline{\epsfxsize=55mm\epsfbox{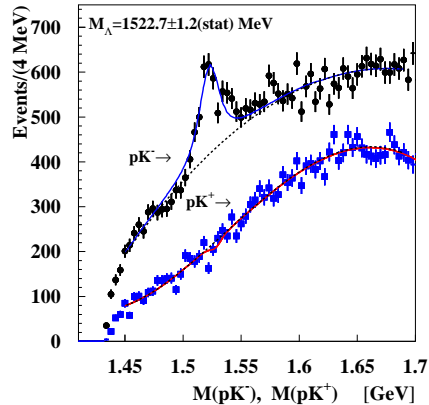}}
\caption{Spectra of invariant mass $M_{pK^-}$ (top) and $M_{pK^+}$
(bottom). A clear peak is seen for the $\Lambda(1520)$ in the
$M_{pK^-}$ invariant mass distribution. However, no peak structure
is seen for the hypothetical $\Theta^{++}$ in the $M_{pK^+}$
invariant mass distribution near 1.53\,GeV. \label{Fig2}}
\end{figure}

Estimates of the spectrometer acceptance times efficiency from the
toy Monte Carlo simulation mentioned above were used to estimate
some cross sections. Taking the branching fraction of
the $\Theta^+$ to $p K^0_S$ to be $1/4$, the cross section for its
photoproduction is found to range from about 100 to 220\,nb $\pm
25\%\mbox{(stat)}$, depending on the model for the background and
the functional form fitted to the peak. The cross section for
photoproduction of the $\Lambda(1520)$ is found to be
$62\pm11\mbox{(stat)}$\,nb.  All of these estimates are subject to
an additional factor of two uncertainty, to account for the
assumptions about the kinematic distribution of the parents used
in the simulation.

A comparison of the mass values reported to date for the $\Theta^+$
state by other experiments to the present results indicates that
there are large variations in mass. However, this is not uncommon
for new decaying particles. Nevertheless, there is clearly a need
for better estimates of experimental uncertainties. By fitting the
available mass values\cite{SPring8}\cdash\cite{LPI} with a constant,
the weighted average is $1532.5\pm2.4$\,MeV. The uncertainty of the
average was scaled by the usual\cite{PDG} factor of square root of
the reduced $\chi^2$.

It is important to note though, that the experimental status of
pentaquark baryons is still controversial. The signal of the \Thp
claimed by Refs.~\refcite{SPring8}--\refcite{LPI} fails to appear in
the data from a large number of other experiments, as shown in
Table~1. Unfortunately, none of these eleven results have been
published at the time of this workshop, and only
three\cite{null_results} have appeared on the e-Print archive
server.\cite{arXiv}
\begin{table}[ht]
\tbl{Summary of null results for three possible pentaquark states.
Evidence for the $ \Xi^{--}(1862)$ and the $\Theta_c(3100)$ has
been reported by Refs.~25 and 26, respectively. Only the null
results from the experiments with a $^*$-symbol have appeared on
the e-print archive server at the time of this workshop.}
{\footnotesize
\begin{tabular}{@{}lcccl@{}}
\hline
{} &{} &{} &{} &{}\\[-1.5ex]
Experiment& $\Theta^{+}$(1540) &  $ \Xi^{--}$(1862)&$\Theta_c(3100)$&Reaction \\
&$(uudd\overline{s})$ &$(ddss\overline{s})$&$(uudd\overline{c})$& \\
{} &{} &{} &{} &{}\\[-1.5ex]
\hline
{} &{} &{} &{} &{}\\[-1.5ex]
HERA-B$^*$& NO &NO & &$p A  \rightarrow \Theta^{+}X,\; \Xi^{--}X$\\
E690 & NO&NO & &$pp  \rightarrow \Theta^{+}X,\; \Xi^{--}X$\\
CDF  &NO& NO& NO & $p\overline{p}  \rightarrow \Theta^{+}X,\; \Xi^{--}X,\; \Theta^{c}X$\\
HyperCP& NO& & &$\pi,K, p \rightarrow \Theta^{+}X$\\
BaBar& NO&NO & &$e^{+}e^{-} \rightarrow \Theta^{+}X,\; \Xi^{--}X$\\
ZEUS&  yes & NO &NO &$ep \rightarrow \Theta^{+}X,\; \Xi^{--}X,\; \Theta^{c}X$\\
ALEPH& NO& NO& NO&$e^{+}e^{-} \rightarrow \Theta^{+}X$\\
DELPHI& NO& & &$e^{+}e^{-} \rightarrow \Sigma ^{-}K^{0}p$\\
PHENIX$^*$ &  NO & & &$AuAu \rightarrow \Theta^{+}X $\\
FOCUS  &   & &NO & $\gamma A \rightarrow \Theta^{c}X$\\
BES$^*$  & NO & &  & $e^+ e^-  \rightarrow J/\Psi \rightarrow \Theta^{+} \overline{\Theta^{-}}$\\
\hline
\end{tabular}\label{table2} }
\end{table}

\subsection{Systematic Studies}

The general experimental situation is quite unsettled at this point.
There are many open questions that need to be answered: how real are
the positive results, and equally, how real are the null results;
what is the actual mass, intrinsic width, the spin and parity, etc.,
of this new particle. In particular, the positive results need to be
checked whether they were produced from fake peaks that can arise
from kinematic reflections or from detector acceptance and kinematic
constraints on the data. While the former check has not been
entirely done for the present result, the \pythia\ and \Thp Monte
Carlos have not produced any fake peaks due to acceptance or
kinematic constraints.

Because the present experiment does not precisely determine the
strangeness of the observed peak, the question has arisen whether
it is a true pentaquark (with strangeness $S$=$+1$) state or a
previously unobserved $\Sigma^{*+}$ resonance. Under the
assumption that the peak is a $\Sigma^{*+}$ resonance, there
should also appear a peak in the $M_{\Lambda \pi^+}$
spectrum.\cite{Poliakov_private} However, as can be seen in
Fig.~3, no peak appears in the $M_{\Lambda \pi^+}$ spectrum near
1.53\,GeV, even though the well established $\Sigma(1385)^+$
baryon resonance is clearly seen. Furthermore, the $M_{\Lambda
\pi^-}$ spectrum clearly shows the well known
$\Sigma(1385)^-$ and $\Xi^-$ states, demonstrating the ability of
the current experiment to identify narrow resonances in $\Lambda
\pi$ invariant mass spectra near 1530\,MeV. This indicates that
the observed peak in the $M_{p\pi^+\pi^-}$ invariant mass spectrum
cannot be a previously unobserved $\Sigma^{*+}$
resonance.\cite{Close_private}

\begin{figure}[ht]
\centerline{\epsfxsize=54mm\epsfbox{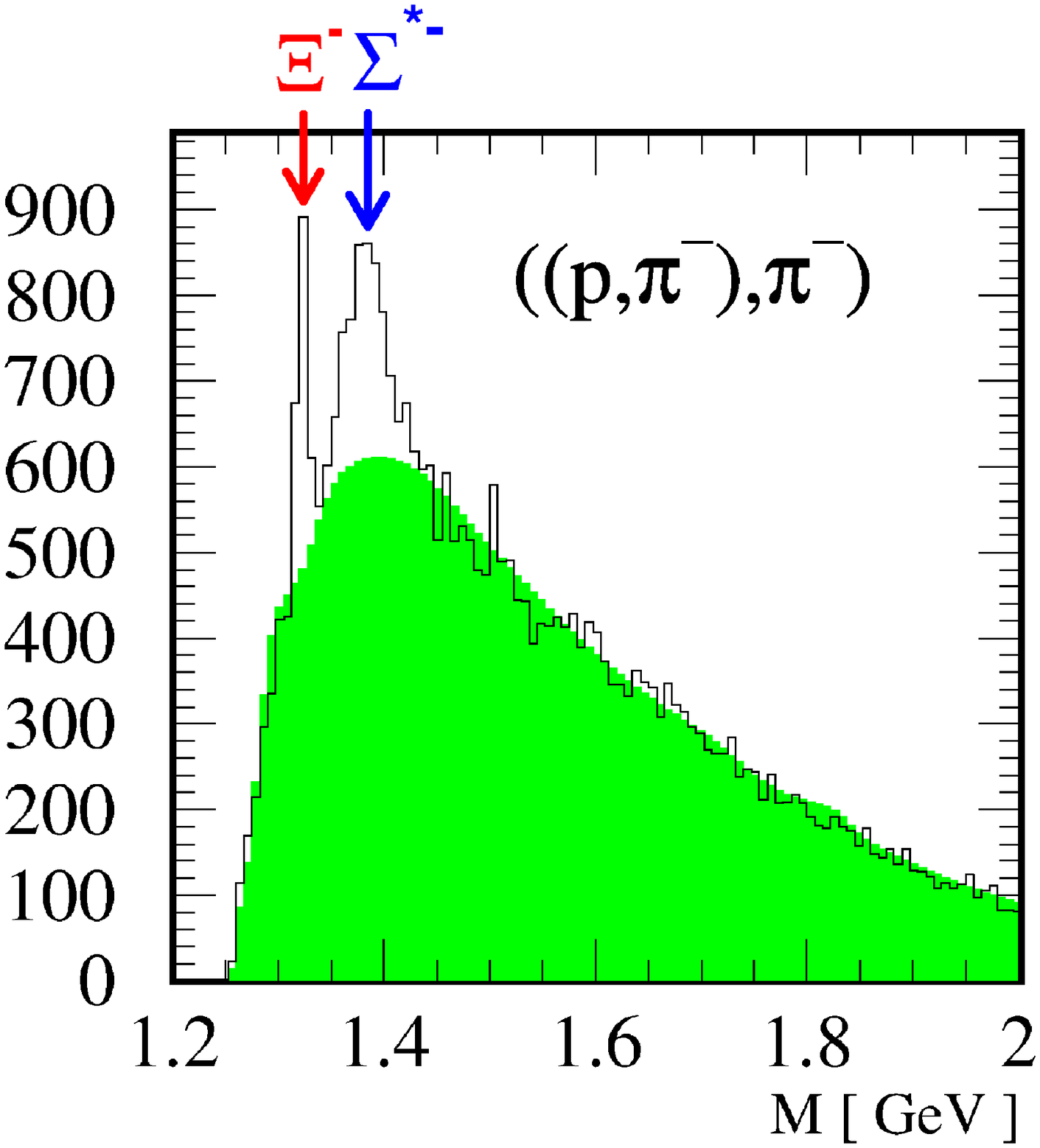}
\hspace*{2mm} \ \epsfxsize=54mm\epsfbox{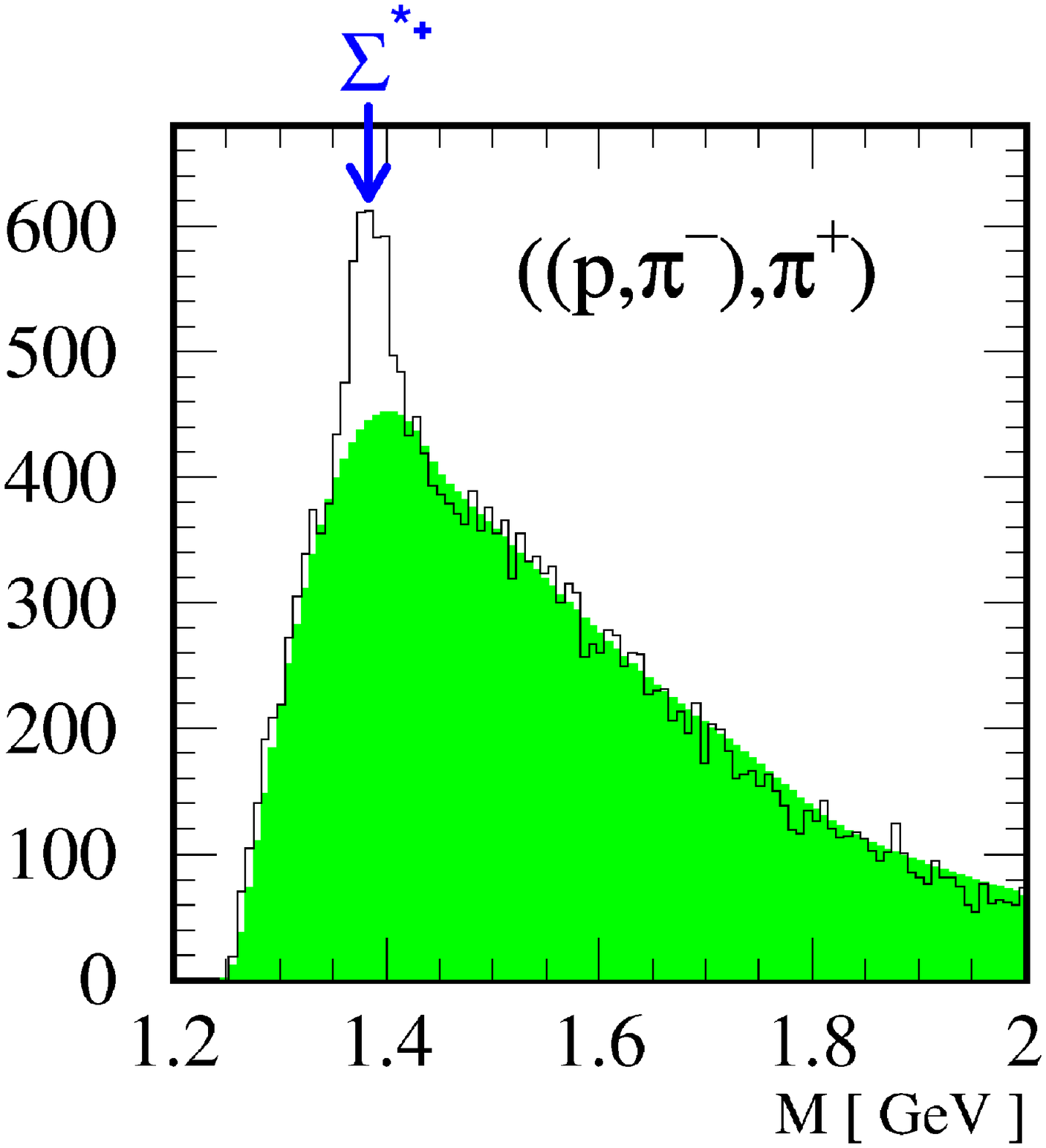}}
\caption{Distribution in invariant mass of the $(p \pi^-)\pi^-$
(left panel) and $(p \pi^-)\pi^+$ (right panel) system. There is
no peak in the $(p \pi^-)\pi^+$ invariant mass spectrum near
1.53\,GeV. The shaded histogram represents the mixed-event
background. \label{Fig3}}
\end{figure}

Although the present experiment has excellent particle
identification for protons and kaons, the $M_{p\pi^+\pi^-}$
invariant mass spectrum exhibits a relatively large background under
the peak of interest. Most of this background is from $K_S$ mesons
that originate from processes other than \Thp decay. A large
fraction of such $K_S$ mesons from exclusive processes can be
removed if an additional hadron is required in the event. Due to the
limited acceptance of the HERMES spectrometer, however, this
additional requirement reduces the number of events that passed all
the kinematic constraints from 1203 to 395,  in 86\% of which the
extra hadron was a pion. Requiring the additional hadron to be a
pion, here termed $\pi_{\rm 4th}$, removes $K_S$ mesons from the
process $\gamma p \rightarrow \phi p \rightarrow K_L^0 K_S^0 p$,
because it is rather unlikely to detect a pion from $K_L$ decay in
our detector. The additional pion further helps to remove $K_S$ from
$p(K^*)^\pm\rightarrow p K_S^0 \pi^\pm_{\rm 4th}$ by introducing a
new veto constraint on the $M_{K_S^0 \pi}$ invariant mass. Since the
proton and the additional pion can come from a $\Lambda\rightarrow p
\pi^-_{\rm 4th}$ decay, an additional veto constraint is placed on
the $M_{p \pi^-_{\rm 4th}}$ invariant mass. Figure~4 shows the \Thp
mass spectrum with an additional pion in the event after applying
these two new constraints (on the $K^*$ and $\Lambda$) in addition
to all the standard kinematic constraints. The mass and the
width of the peak are in good agreement with the published
results,\cite{HERMES} however, the ratio of signal to background
improves from 1:3 (see Fig.~1) to 2:1.

\begin{figure}[ht]
\centerline{\epsfxsize=55mm\epsfbox{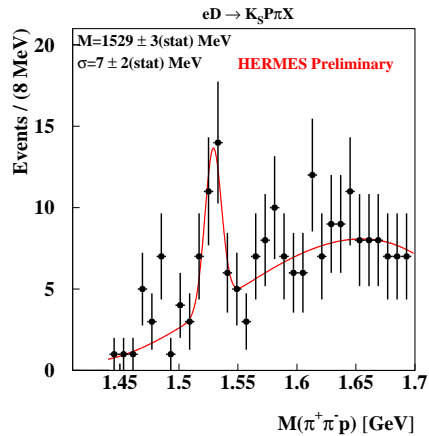}} \caption{Invariant
mass distribution of $M_{p\pi^+\pi^-}$ with an additional pion,
subject to the constraints in event topology discussed in the text.
\label{Fig4}}
\end{figure}

It was further investigated whether the fourth hadron could come
from the following exclusive processes:  $\gamma p \rightarrow
 \overline{K^0}\Theta^+ \rightarrow
(\pi^+ \pi^-)(K_S^0 p \rightarrow \pi^+ \pi^- p)$, or $\gamma n
 \rightarrow K^-\Theta^+ \rightarrow K^-(K_S^0
p \rightarrow \pi^+ \pi^- p)$. Results from a Monte Carlo study
revealed that the associated $K^-$ or $K_S$ from these exclusive
processes go to backward angles, and that even the pions from $K_S$
decay are inaccessible with the HERMES detector. This is due to the
PID threshold on the proton, which requires that the momentum of the
\Thp must be larger than 7\,GeV. Therefore, the tagged pion events
cannot originate from these exclusive processes, which implies that
the production cross section has to be at least partially inclusive.
This is an interesting observation because it does not support a
possible explanation for the discrepancy between the positive
results from low energy experiments, which are mainly from exclusive
reactions, and the null results at high energy experiments, which
are primarily inclusive measurements.  The idea was that the
$\Theta^+$ would be produced in exclusive processes, and the cross
sections for such processes typically decrease with increasing
energy (with the exception of elastic scattering).  Even though the
cross sections for inclusive processes tend to increase with energy,
it was hypothesized that the $\Theta^+$ would not be produced in
inclusive processes, thus failing to appear in high energy
experiments. However, the present new data (as well as the
observation of the $\Theta^+$ by ZEUS\cite{ZEUS}) appear to
contradict this idea.

\section*{Acknowledgments}

I wish to thank my colleagues in the HERMES collaboration. I
acknowledge Avetik Airapetian and Andy Miller for critical reading
of the manuscript. The author's research is supported in part by the
U.S. National Science Foundation, Intermediate Energy Nuclear
Science Division under grant No. PHY-0244842.

\end{document}